\documentclass[aps,prl,amsmath,amssymb,reprint,balancelastpage,floatfix,showpacs,showkeys,footinbib]{revtex4-1}

\usepackage{graphicx}% Include figure files
\usepackage{dcolumn}% Align table columns on decimal point
\usepackage{bm}% bold math
\usepackage{subfigure}
\usepackage{xcolor}
\usepackage{times}

\newcommand{\unit}[1]{\ensuremath{\, \mathrm{#1}}}
\newcommand{\um}{\ensuremath{\,\mu \mathrm{m}}}

\begin{document}

\title{Electric field sensing near the surface microstructure of an atom chip using cold Rydberg atoms}
\author{J.~D.~Carter}
\author{O.~Cherry}
\author{J.~D.~D.~Martin}
\affiliation{Department of Physics and Astronomy and Institute for Quantum Computing, \\ University of Waterloo, Waterloo ON, N2L 3G1, Canada}

\date{\today}% It is always \today, today,

\begin{abstract}
The electric fields near the heterogeneous metal/dielectric surface of an atom chip were measured using cold atoms. The atomic sensitivity to electric fields was enhanced by exciting the atoms to Rydberg states that are 10$^8$ times more polarizable than the ground state.  We attribute the measured fields to charging of the insulators between the atom chip wires. Surprisingly, it is observed that these fields may be dramatically lowered with appropriate voltage biasing, suggesting configurations for the future development of hybrid quantum systems.
\end{abstract}

\pacs{ % http://www.aip.org/pacs/pacs2010/individuals/pacs2010_regular_edition/index.html
32.80.Ee, % 	Rydberg states 
34.35.+a, % Interactions of atoms and molecules with surfaces 
37.10.Gh % Atom traps and guides 
}% PacS, the Physics and Astronomy Classification Scheme.
\keywords{Rydberg atoms, atom chips}%Use showkeys class option if keyword
\maketitle

Hybrid quantum systems seek to combine the benefits of gas-phase ultra-cold atoms or molecules (long coherence times for information storage) and solid-state quantum devices (strong interactions for fast gates) \cite{PhysRevLett.92.063601,hogan:2011,andre:2006}. Rydberg, or ``swollen'', atoms -- atoms with a highly excited valence electron -- may enable hybrid devices by amplifying the interactions between atoms and devices in a similar manner to the enhancement of interactions between atoms \cite{isenhower:2010,gaetan:2009}.  However, these hybrid systems require atoms to be located near a heterogeneous surface with exposed metal electrodes and dielectric insulators, which can be sources of uncontrollable and unwanted electric fields.  

\begin{figure}[!]
\vspace{0.1in}
\centering
\subfigure{(a)
\includegraphics[width=3.375in]{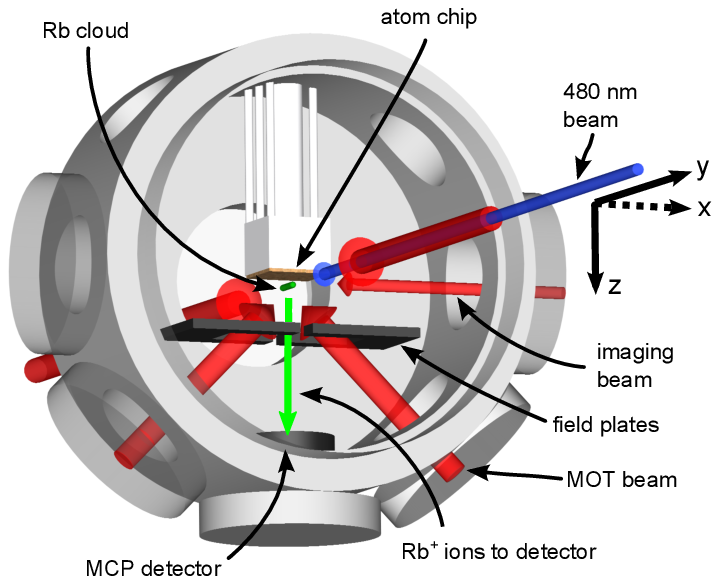}} \\
\vspace{0.1in}
\subfigure{(b)
\includegraphics[width=1.6in]{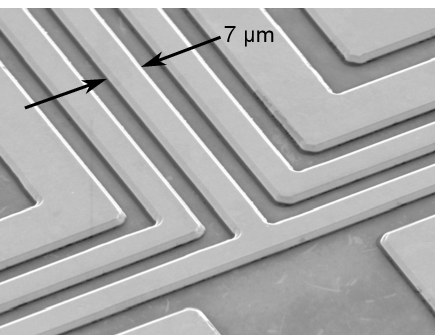}} 
\subfigure{(c)
\includegraphics[width=1.3in]{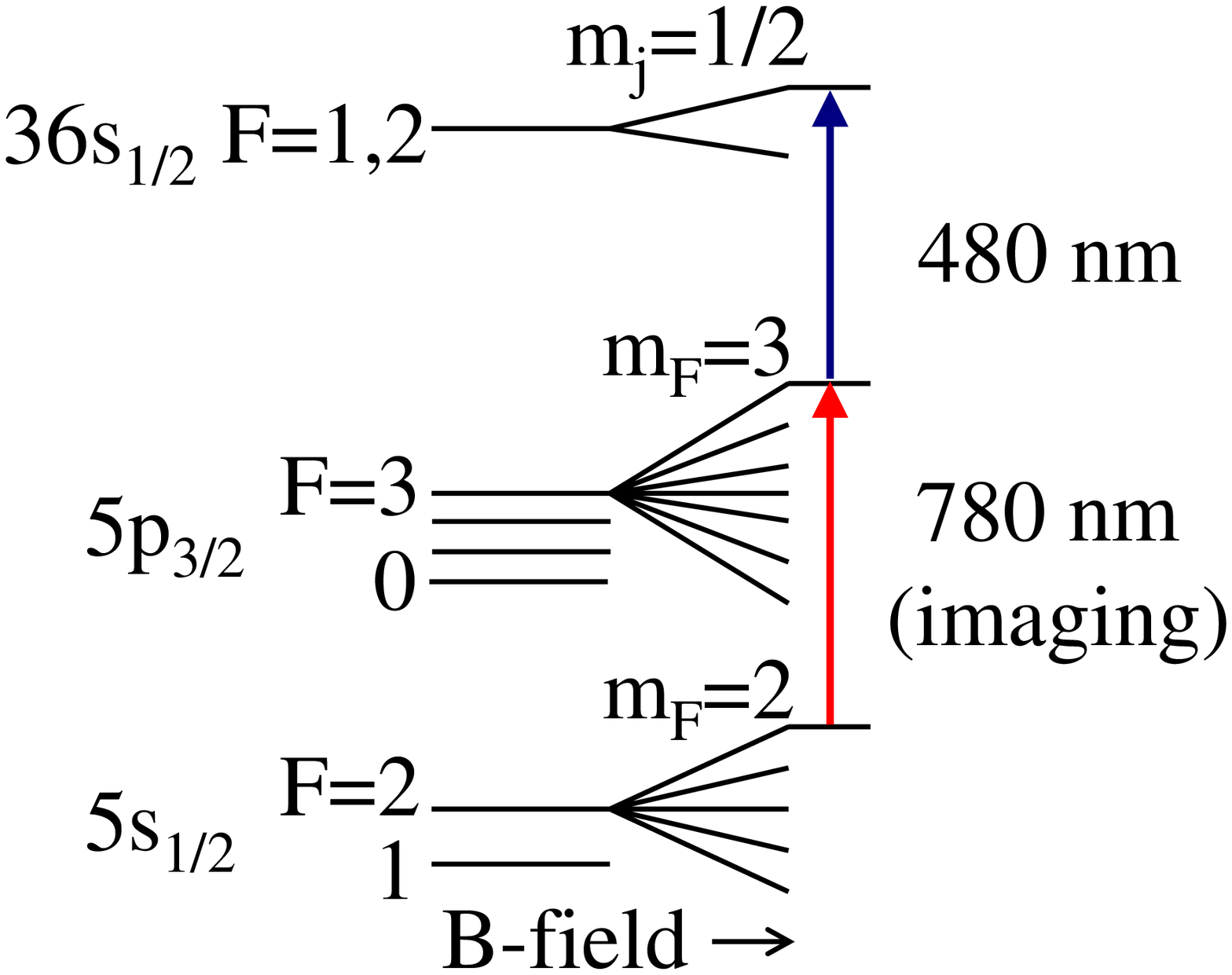}} \\ \vspace{0.2in} 
\subfigure{(d)
\includegraphics[width=3in]{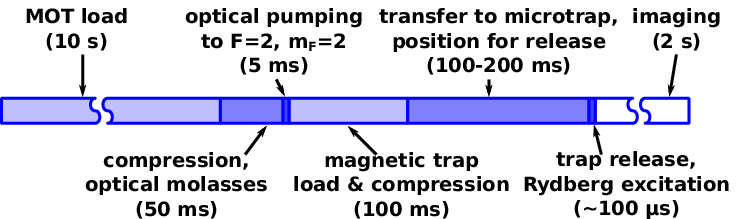}}
\caption{
(a) Experimental apparatus. (b) Scanning electron microscope image of the atom chip at one end of trapping region, showing wires and insulating gaps.
(c) $^{87}$Rb Rydberg excitation scheme (see for example Ref.~\onlinecite{low:2012}). 
(d) Experimental sequence timing.  A single cycle takes 
$\approx 15\unit{s}$.
\label{fg:apparatus}}
\end{figure}

Rydberg atoms have a high susceptibility to small electric fields \cite{gallagher:1994}  and this can be problematic near surfaces.
For example, to study the intrinsic ``image-field'' ionization of Rydberg atoms near a metal surface one must avoid adsorption of contaminants and use flat, single-crystal orientation surfaces \cite{hill:2000,pu:2010}.  Even flat polycrystalline metal surfaces may generate significant inhomogeneous electric fields due to the differing work-function between grains \cite{carter:2011}. In addition to static fields, surfaces may also be a source of enhanced fluctuating fields: a problem which plagues ion-trapping (see Ref.~\onlinecite{low:2011} and references therein) and is also a consideration for Rydberg atoms near surfaces \cite{muller:2011}.
For dielectrics, which are a necessary part of any non-trivial device --- as insulating gaps for instance --- charging and time-dependent electric fields due to adsorbates \cite{abel:2011} must also be considered.

Atom chips \cite{fortagh:2007,reichel2010atom} offer the ability to trap cold neutral atoms close to surfaces, and observe the influence of surfaces \cite{lin:2004}. This technology has recently been exploited by Tauschinsky {\it et al.}~\cite{PhysRevA.81.063411} to study the shifts of Rydberg states due to adsorbates on metal surfaces as a function of distance away from a metal surface (a shield between the chip wires and atoms).

In this work, we describe experiments incorporating laser cooled $^{87}$Rb, an atom chip, Rydberg excitation, and charged particle detection (see Fig.~\ref{fg:apparatus}).  This allows the sensing of electric fields near atom chip wire structures, with insulating gaps between wires that are typical of surface devices.  Although the Stark effect is well-known and exploited (in plasma diagnostics for example) this is the first work in which the Stark effect has been used to probe unknown fields near a \mbox{microstructured} surface.

The experimental sequence is shown in Fig.~\ref{fg:apparatus}:
atoms are first loaded from background $^{87}$Rb vapor into a mirror magneto-optical trap (MOT) \cite{wildermuth:2004}, compressed, and optically pumped into the $5s_{1/2},F=2,m_F=2$ sublevel.  The atoms are trapped by quickly turning on a mm-scale magnetic trap and then adiabatically transferred to the trapping potential formed by the atom chip wires.  The atom chip consists of $1\unit{\mu m}$ tall gold wires deposited on a thin $20\unit{nm}$ layer of insulating silicon oxide on a silicon substrate \cite{cherry:2009}. There are five wires on the chip surface: a central H-shaped structure (connected so that the current runs in a z-shape), and two pairs of nested U-shaped wires. In the $4 \unit{mm}$ long trapping region, the wires are arranged closely to each other and run parallel.   The three innermost wires are $7 \unit{\mu m}$ wide and the outer wires are $14 \um$ wide. All wires are separated by gaps of $7 \um$.  The remainder of the $2\times2\unit{cm}$ square chip is covered with a grounded $1\unit{\mu m}$ layer of gold.  In this work, the potential minimum is located between $35-70 \unit{\mu m}$ from the surface of the chip. 

We do not magnetically trap Rydberg atoms \cite{choi:2005} --
the atoms are released from the microtrap prior to Rydberg excitation, because inhomogeneous magnetic fields (due to wire currents) and electric fields (due to the associated voltage drops) broaden the transition and reduce the available signal level.  

After a variable chip hold time, the cloud is adiabatically moved to the desired distance from the chip surface and the chip wires are then quickly shut off.  Rydberg excitation is done $30 \unit{\mu s}$ later, after fields due to eddy currents associated with the wire shutoff have dissipated.  After the chip wires are shut off, a homogeneous magnetic field of $34.5\unit{G}$ remains in the $x$-direction (the microtrap ``bias field'').  A $30 \unit{\mu s}$ long optical pulse excites Rydberg atoms via a two-step process: 1) a $\approx 780\unit{nm}$ laser tuned to the $5s_{1/2},F=2,m_{F}=2 \rightarrow 5p_{3/2},F=3,m_{F}=3$ transition, and 2) $\approx 480 \unit{nm}$ laser light to drive the $5p_{3/2}, F=3, m_{F}=3 \rightarrow 36s_{1/2}$ transition.  We study excitation to Rydberg states after release from the microtrap, varying distance by moving the $480\unit{nm}$ beam relative to the surface using servo-actuated mirrors (staying parallel to the surface).  Some locations are far from the center of the microtrap, but where the atom density is still sufficiently high.

\begin{figure}[tb]
\centering
\includegraphics[width=3in]{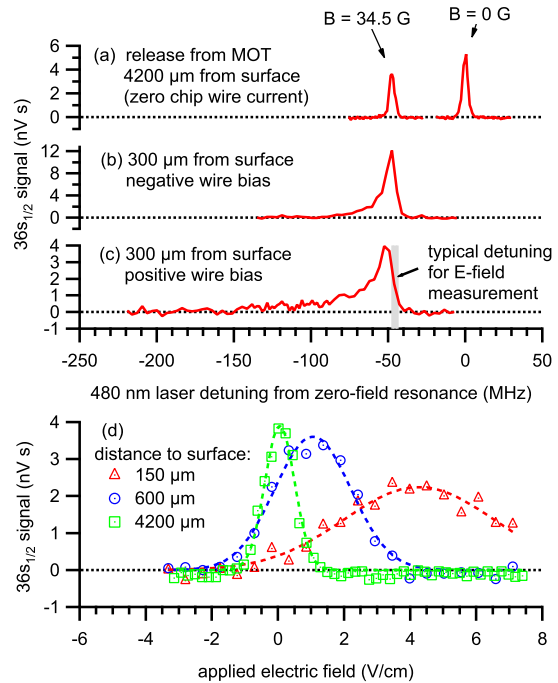}

\caption{
Rydberg excitation spectra after release from the (a) MOT and (b-c) microtrap, with compensating fields applied (see text for details of compensation and wire biasing). (d) Measurement of the Rydberg signal as a function of applied electric field
(by varying plate voltages, corrected for MCP fringing field and ac line interference), with Gaussian fits.  Positive wire bias (see text) was used for the spectra at $150\unit{\mu m}$ and $600\unit{\mu m}$, whereas the result for the larger distance $4200\unit{\mu m}$ was obtained by release from the MOT.
\label{fg:spectra}}
\end{figure}

The Rydberg atoms are detected by selective field ionization (SFI) \cite{gallagher:1994}:  a slowly rising ($\approx \unit{\mu s}$) negative voltage pulse is applied to the two metal plates away from the chip surface (see Fig.~\ref{fg:apparatus}), creating a field normal to the chip surface. 
Ionized Rb atoms are drawn towards a microchannel plate (MCP) detector.
Optical spectra for excitation of the $36s_{1/2}$ state are shown in Fig.~\ref{fg:spectra}(a)-(c).  Far from the surface the linewidths are narrow, roughly dictated by the $5p_{3/2}$ radiative lifetime.
When the atoms are within about $300 \unit{\mu m}$ from the surface, the optical spectra broaden and become asymmetric. Both effects are caused by Stark shifting due to inhomogeneous electric fields --- the $36s_{1/2}$ level shifts quadratically towards lower energy as the field $F$ increases \cite{gallagher:1994}: $\Delta E = -(\alpha/2) F^2$ with $\alpha/2 \approx 2.6 \unit{MHz/(V/cm)}^2$.  For comparison, $\alpha/2 \approx 0.04 \unit{Hz/(V/cm)}^2$ for the ground state of Rb.

\begin{figure}[tb]
\centering
(a)
\includegraphics[width=3in]{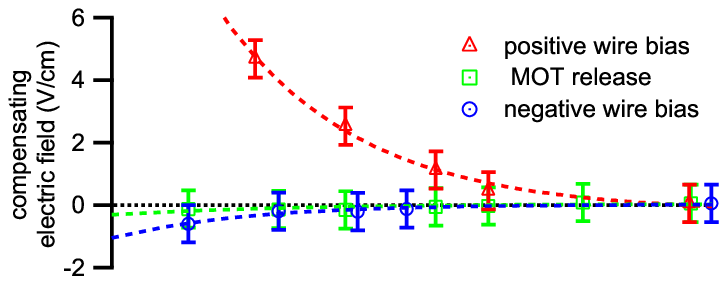}\\
(b)
\includegraphics[width=3in]{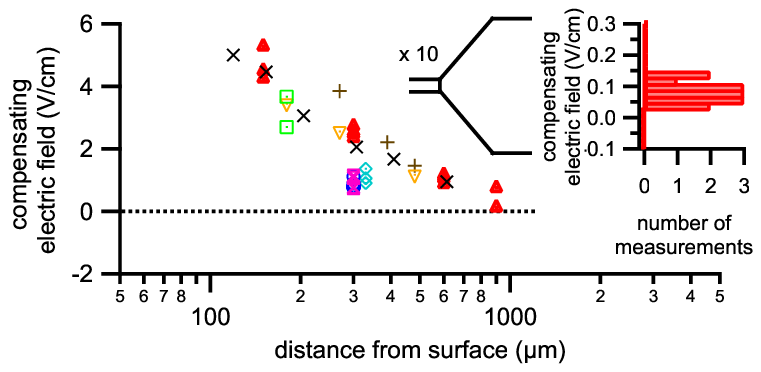}\\
(c)
\includegraphics[width=3in]{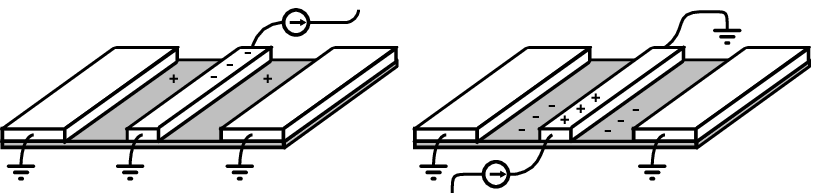}
\caption{
(a) Distance dependence of compensating field (corrected for MCP fringing fields and ac line interference; see Supplementary Materials) measured after microtrap release and MOT release, with power-law fits. (b) Compensating field measured at various distances from the chip, taken over 8 different days (represented by different point styles) in a two-month period.   In all microtrap measurements in (a)-(b), the atoms were held in the trap for 225 ms prior to release. Inset: histogram of 14 compensating field measurements $3 \unit{mm}$ from the surface, taken after release from the MOT, on 13 days in a two-month period. (c) Charge accumulation in the dielectric gaps near a negatively biased wire (left) and positively biased wire (right).
\label{fg:fieldsdistances}}
\end{figure}

The $36s_{1/2}$ state is red-shifted by electric fields. Therefore, we measure the ``average'' normal electric field component by blue-detuning the Rydberg excitation laser about half a linewidth from resonance (as illustrated in Fig.~\ref{fg:spectra}(c)) and varying an applied electric field created by biasing the field plates.  Figure \ref{fg:spectra}(d) shows signal vs.~applied field at three distances from the surface. The signal is maximized when the applied electric field cancels the ``average'' electric field near the chip (the fields near the chip are inhomogenous so this cancellation will not be complete for all locations). This value of the applied field is called the ``compensating field''. For a given distance, we determine the compensating field from the center of a fitted Gaussian.  The spectra in Fig.~\ref{fg:spectra}(a)-(c) are taken with compensating fields applied.

We observe that the voltages of the chip wires in the microtrapping phase significantly affect the electric fields near the chip.  This is surprising, as the currents are turned off and the wires grounded prior to Rydberg excitation.   However, there is no significant time dependence when measuring the field at several different times ranging from $30-100 \unit{\mu s}$ after release from the microtrap.  For typical operating currents, the electrical resistance of a chip wire causes a potential drop of about $6 \unit{V}$ along its length. Since the current supply holds one end of the wire near ground, the wire will have an overall biasing of several volts relative to ground. This biasing varies along the wire's length and can be positive or negative, depending on whether the supply sources or sinks current.  We refer to these conditions as ``positive'' or ``negative wire bias''.  Spectra obtained when the chip wires were positively biased consistently show more broadening and lower signal levels compared to negative biasing, even though the magnetic field geometry is identical. 

The distance dependence of the measured average compensating fields are plotted in Fig.~\ref{fg:fieldsdistances}. 
When atoms are released from the microtrap, the scaling of the measured field with distance is consistent with a $1/z$ power law, with fitted power-law scalings of $z^{-0.99\pm0.3}$ and $z^{-0.93\pm0.1}$ for negative and positive wire potentials, respectively. The electric field direction depends on the wire biasing, consistent with a positive surface charge when the wire potential is negative and vice versa. The plot in Fig.~\ref{fg:fieldsdistances}(b) illustrates the day-to-day measurement repeatability. Measurements taken far from the chip (where the fields are homogeneous and do not vary from day to day) are quite consistent, reproducible to within $0.04 \unit{V/cm}$. Our estimate of the measurement uncertainty due to detection signal/noise \footnote{See Supplemental Materials at end for additional experimental details and measurement sensitivity.} 
is consistent with this reproducibility.  Therefore, most of the variability in field measurements made close to the surface is in fact due to changes in the fields from day to day. 
\begin{figure}[tb]
\centering
\includegraphics[width=3in]{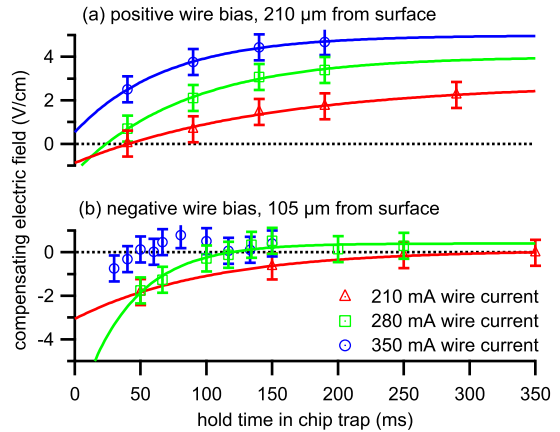}

\caption{
Compensating field as a function of hold time in microtrap together with exponential fits, (a) positive wire bias, Rydberg excitation $210 \unit{\mu m}$ from surface. (b) negative wire bias, Rydberg excitation $105 \unit{\mu m}$ from surface. 
\label{fg:fieldstime}}
\end{figure}

A possible explanation for the observed distance scaling and direction of the field is that ambient charged particles are drawn toward oppositely-biased wires, as illustrated in Fig.~\ref{fg:fieldsdistances}(c), and then trapped in the insulating gaps between the wires.  They remain there for some time ($>1\unit{ms}$) even after the chip wires are shut off and the wires are at ground (consistent with the observed lack of time-dependence of the fields after release on the $100\unit{\mu s}$ timescale).  Such a charging mechanism should saturate.  This is seen for positively biased wires in Fig.~\ref{fg:fieldstime}(a), where the field magnitude depends exponentially on the amount of time the microtrap wires are turned on before the atoms are released, and at long times approaches a value proportional to the wire current (and thus the biasing potential). 

In this explanation there is a natural asymmetry between the positive and negative biasing cases due to the differing mobilities and trapping of oppositely signed charges.  When the wires are negatively biased, we do not observe charge accumulating over time. Instead, the gaps appear to have a significant net positive charge shortly after the wires are turned on, and the charge neutralizes as the wires operate.  The rate of neutralization depends strongly on wire current (see Fig.~\ref{fg:fieldstime}), suggesting a thermally activated neutralization mechanism. Charge transfer between the dielectric surface and the semiconducting substrate --- which is in contact with current-carrying metal structures below the chip --- is one possible explanation for the initial charging in this case. 

When atoms are released from the MOT, rather than the microtrap, the measured field direction is consistent with a small positive charge on the surface.  However, the magnitude is smaller than when atoms are released from the microtrap, and has a weaker distance dependence,with $1/z^{0.67\pm 0.2}$ scaling. Turning on the chip wires while the atoms are trapped in the MOT (rather than the microtrap) has no effect on the measured electric field after release, a result which is inconsistent with a slowly-relaxing dielectric polarization as an explanation for the fields \cite{abel:2011}.

The field direction in the negative wire bias case is consistent with Rb deposited preferentially near the center of the chip \cite{PhysRevA.75.062903}, and  the distance scaling we observe is similar to Ref.~\onlinecite{PhysRevA.81.063411}. However, the fields we observe are an order of magnitude smaller and do not change when we deposit Rb on the surface by deliberately moving the cloud close to the chip. This diminished adsorbate field is encouraging for the study of intrinsic Rydberg atom surface phenomena, such as the Lennard-Jones shift \cite{sandoghdar:1996} (using chips with an electrostatic shield between the wires and atoms \cite{PhysRevA.81.063411}).

The quantitative behavior we have observed is for a specific geometry, but our measurement approach and the influence of biasing are quite general.  
For example, recent experiments by Hogan {\it et al.}~\cite{hogan:2011} involving Rydberg atoms close to a co-planar waveguide may also benefit from this type of dc biasing (inner-conductor negative with respect to ground).
Although we have exploited the high sensitivity of Rydberg atoms to measure electric fields, cold ground-state atoms  \cite{kruger:2003,PhysRevA.75.062903} and molecules \cite{andre:2006} also exhibit sensitivity to electric fields, and similar biasing considerations apply.

In summary, we have performed Rydberg atom sensing of electric fields near a microstructure consisting of gold wires and insulating gaps.  We have observed an electric field due to charging --- however, voltage biasing of the chip wires with respect to the surrounding grounded surfaces can dramatically reduce this charging.    In the future, our demonstration of selective-field-ionization near the chip can be extended to state-sensitive detection of Rydberg atoms, enabling the use of microwave transitions between Rydberg states for noise spectroscopy \cite{bylander:2011} near the chip surface.  This would establish limits on the coherent manipulation of Rydberg atoms near atom chips due to electric field noise \cite{muller:2011} and help test surface noise models \cite{safavi:2011}.

\begin{acknowledgments}
We thank R.~Mansour for use of the CIRFE facilities, J.~B.~Kycia for the loan of equipment, and C.~E.~Liekhus-Schmaltz for comments on this manuscript.  This work was supported by NSERC.
\end{acknowledgments}

\renewcommand{\figurename}{Supplementary Figure}
\setcounter{figure}{0}

\newpage

\noindent
{\large Supplementary Material for ``Electric field sensing near the surface microstructure of an atom chip using cold Rydberg atoms'',
 Carter {\it et al.}}

\section{Experimental details}

\noindent
{\bf Trap loading:}
We now give more specific details concerning the experimental sequence shown in Fig.~1 of the main text:
Atoms are first loaded from background $^{87}$Rb vapor (supplied with dispensers) into a mirror magneto-optical trap (MOT) centered 2-3 mm below the chip surface.   The quadrupole field is generated by a current-carrying U-shaped structure underneath the chip and external field coils.  Typically $10-20 \times 10^6 \unit{atoms}$ are loaded in about $10 \unit{s}$. 

The cloud is then compressed by increasing the cooling laser detuning to reduce the radiation pressure.
After compression, the quadrupole field is ramped down, with the MOT beams left on to slow the expansion of the cloud and damp any acceleration due to transient magnetic field gradients caused by eddy currents.
The MOT beams are then turned off and the atoms are optically pumped into the weak field-seeking $F=2,m_F=2$ sublevel.  The atoms are then confined by quickly turning on a mm-scale magnetic trap formed by a current-carrying z-shaped structure below the chip and external field coils.  More than $2/3$ of the MOT population can be successfully captured in the magnetic trap.  The $1/e$ lifetime of the cloud in this trap is typically $2-4 \unit{s}$, consistent with the loss rate due to collisions with room-temperature background gas at a pressure of $10^{-9} \unit{Torr}$.  The cloud is adiabatically transferred to the microtrap by ramping up the current in the chip wires and then slowly ramping down the current in the larger wire below the chip.  There is some atom loss due to evaporation in this process.  The initial population of the chip trap is about $1.5 \times 10^6$ and decays exponentially with a time constant of around 500 ms. 

\noindent
{\bf Atom chip:}
The atom chip consists of $1\unit{\mu m}$ high gold wires deposited on a thin $20\unit{nm}$ layer of insulating silicon oxide on a silicon substrate.  There are five wires on the chip surface: a central H-shaped structure (connected so that the current runs in a z-shape), and two pairs of nested U-shaped wires.
In the $4 \unit{mm}$ long trapping region, the wires are arranged closely to each other and run parallel.   
The three innermost wires are $7 \unit{\mu m}$ wide and the outer wires are $14 \um$ wide. All wires are separated by gaps of $7 \um$. The potential created by wire currents and external magnetic field coils has approximate cylindrical symmetry, though field gradients are largest near the chip surface.  Details of the fabrication of the atom chip are contained in Cherry {\it et al.}, Can. J. Phys. {\bf 87} 633 (2009) (see Fig.~3 in this reference for the exact wire geometry).

\noindent
{\bf Optical excitation:}
The $780\unit{nm}$ light for cooling and trapping is produced by two external-cavity diode lasers.   The $480\unit{nm}$ light for Rydberg excitation is obtained by frequency doubling a Ti:sapphire laser that is stabilized using a transfer cavity; see C. E. Liekhus-Schmaltz {\it et al.} 
J. Opt. Soc. Am. B, {\bf 29}, 1394 (2012). 

During Rydberg excitation, the $780\unit{nm}$ light is introduced in the same way as for absorption imaging (along the $x$-axis; see Fig.~1), whereas the $480\unit{nm}$ light travels along the long $y$-dimension of the released cloud, with vertical polarization ($z$-direction).  The $480\unit{nm}$ light has a beam waist of $w=30\unit{\mu m}$ ($1/e$ amplitude radius), and a Rayleigh range of $z_R=5\unit{mm}$ (measured using a scanning knife edge).  

In this work, the two-photon Rydberg excitation is resonant with the intermediate $5p_{3/2}$ state. However, the observed linewidth of the Rydberg excitation is slightly narrower than predicted by the natural linewidth of the $5p_{3/2}$ state.
By releasing atoms from the MOT and then performing Rydberg excitation at distances far from the chip ($4.2\unit{mm}$), we observe a linewidth of $3.6\pm0.2\unit{MHz}$ (see Fig.~2(a)) instead of the expected $6.0 \unit{MHz}$. This result was found in both zero magnetic field and in a homogeneous magnetic field of the same magnitude as the microtrap bias field. A possible explanation is coherence narrowing due to radiation trapping; see
M.~I.~D’yakanov and V.~I.~Perel, Soviet Phys. JETP {\bf 20}, 997 (1965).

\noindent
{\bf Measurement of the electric fields:} The Stark shift of the $36s_{1/2}\rightarrow36p_{1/2}$ microwave transition was used to calibrate the applied compensating electric field in terms of field plate bias voltage. This technique was also used to measure fringing fields from the front of the MCP detector (normally held at $-1800 \unit{V}$ relative to ground, but varied to determine its contribution to the field near the chip). This microwave transition has the advantages of narrower linewidth and a higher electric field sensitivity compared to the optical $5p_{3/2}\rightarrow 36s_{1/2}$ transition. 

In addition to correcting for the fringing field ($1.88 \pm 0.09 \unit{V/cm}$), we also corrected variations in the measured electric field due to slowly time-varying fields associated with the ac line --- the measured field varies sinusoidally at the ac electrical power line frequency, with an amplitude of $0.24 \unit{V/cm}$.  Neglecting the effects of Rb adsorption, we would expect to see a small dc field on the order of $0.1 \unit{V/cm}$ due to the work function difference between the gold chip surface and the stainless steel field plates, but measurements taken far from the chip surface are consistent with zero field once the above corrections have been applied.

Measurements of the electric field taken $3 \unit{mm}$ from the surface, where the effects of charging will be very weak, are reproducible to within $0.04 \unit{V/cm}$, as shown in the inset of Fig.~3(b). Measurements taken on the same day under nominally identical conditions are reproducible to within $0.15 \unit{V/cm}$, but the day-to-day variability is larger. The data shown in Fig.~3(b) are consistent with an overall measurement uncertainty of $0.6 \unit {V/cm}$. We are currently working to identify the sources and conditions influencing this variability.

\section{Electric field measurement uncertainty}
\begin{figure}[b]
\centering
\includegraphics[width=3in]{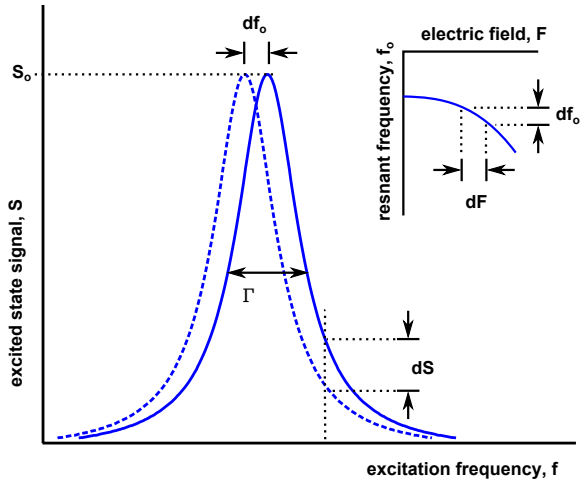}
\caption{
A small change $dF$ in the electric field shifts the transition by some amount $df_o$, via the Stark shift. This changes the measured signal level by $dS$.
\label{fg:resonance}}
\end{figure}

Fluctuations in the detected signal limit the precision of electric field measurements as follows.
Consider an atomic transition, with maximum signal $S_o$ at the resonant frequency $f_o$, and a linewidth $\Gamma$, with the atoms in an electric field $F$.
If the electric field is changed by a small amount, as illustrated in Supp. Fig. \ref{fg:resonance}, the Stark shift changes the transition energy. Therefore, the observed signal will change (with the excitation frequency kept constant) according to 
\begin{equation}
\frac{dS}{dF}=\left(\frac{dS}{df_o}\right)\cdot\left(\frac{df_o}{dF}\right),
\label{eq:derivative}
\end{equation}
where the first factor is determined by the line shape and detuning of the excitation frequency from resonance, and the second factor by the Stark shift. 
If a single measurement of the the excited state signal has some uncertainty $\delta_S$ (perhaps due to detector noise), then the measurement of the local field (by varying the compensating field) has an uncertainty on the order of
\begin{equation}
\delta_F\approx\frac{\delta_S}{(\unit{d}S/\unit{d}F)\sqrt{N}}, 
\end{equation}
where $N$ is the number of measurements. Therefore, maximum measurement precision occurs under conditions where $(\unit{d}S/\unit{d}F)$ is maximum.

If the line shape is Lorentzian, the maximum possible magnitude for the first factor in Eq.~\ref{eq:derivative} is
\begin{equation}
\frac{dS}{df_o}=\frac{1.30 S_o}{\Gamma},
\end{equation}
when the excitation frequency is detuned by $\Gamma/(2\sqrt{3})\approx 0.29\Gamma$ from resonance.
The numerical factor depends only slightly on the line shape--for example, if the line shape is Gaussian (perhaps because of broadening in an inhomogeneous field) then the numerical factor is 1.43.

If the Stark shift is quadratic, $\Delta E=-(\alpha/2)F^2$ and the maximum possible precision of the field measurement (with optimal excitation frequency) for a given set of experimental conditions is
\begin{equation}
\delta_F\approx\frac{\Gamma}{1.30 S_o \alpha F}\frac{\delta_S}{\sqrt{N}}.
\label{eq:homogeneous_ef}
\end{equation}
This result is useful for estimating the measurement uncertainty in situations where the applied field and linewidth are both known. 

In addition, Eq.~\ref{eq:homogeneous_ef} qualitatively shows how the measurement precision can be improved by increasing the field and using highly polarizable states with long lifetimes. However, if the field is not completely homogeneous, the transition will start to broaden as the polarizability and applied field increase. Therefore, the linewidth $\Gamma$ and maximum signal $S_o$ depend on the polarizability, applied field, and field inhomogeneity.

To estimate the ultimately achievable precision, the effects of field inhomogeneities must be considered. Due to the Stark effect, a field inhomogeneity $\Delta F$ will cause an additional contribution to the linewidth, given by
\begin{equation}
(\Delta\Gamma)=\alpha F(\Delta F)+\frac{\alpha}{2}(\Delta F)^2.
\label{eq:broadening}
\end{equation}
The second term is important only for large field inhomogeneity, such that transition is significantly broadened when the average field $F$ is zero. If we assume that this broadening adds in quadrature with $\gamma$, the linewidth in the limit of highly homogeneous field, then
\begin{equation}
\Gamma^2=\gamma^2+(\Delta\Gamma)^2.
\end{equation}
This additional broadening also shifts some of the population out of resonance with the excitation, reducing $S_o$:
\begin{equation}
S_o=\frac{S_H\gamma}{\Gamma},
\end{equation}
where $S_{H}$ is the maximum signal when the field is highly homogeneous.

Explicitly including the effects of the field inhomogeneity, we modify Eq.~\ref{eq:homogeneous_ef}:
\begin{equation}
\delta_F=\frac{\gamma^2+(\Delta\Gamma)^2}{1.30\gamma\alpha F}\cdot\frac{\delta_S}{S_H\sqrt{N}}.
\end{equation}
The minimum uncertainty for a given polarizability $\alpha$ and field inhomogeneity $\Delta_F$ is found by optimizing the applied field $F$. 

In the limit of small inhomogeneity, $\alpha(\Delta F)^2\ll\gamma$, the first term in Eq. \ref{eq:broadening} dominates, and the minimum uncertainty is
\begin{equation}
\delta_F=\frac{2(\Delta F)}{1.30}\cdot\frac{\delta_S}{S_{H}\sqrt{N}}.
\label{eq:ultimate_limit}
\end{equation}
In this limit, the optimal field is $F=\gamma/\alpha\Delta F$, at which point the broadening due to field inhomogeneity is equal to the natural linewidth, i.e., $\Delta \Gamma=\gamma$. This ultimate limit is independent of $\gamma$ and $\alpha$. However, narrow linewidth and large polarizability allow the condition for maximum sensitivity to be achieved with reasonably small applied field.

If the field inhomogeneity is large, such that$\alpha(\Delta F)^2\gg\gamma$, the second term in Eq. \ref{eq:broadening} dominates. The minimum uncertainty achievable in these conditions is
\begin{equation}
\delta_F=\frac{2(\Delta F)}{1.30}\cdot\frac{\alpha(\Delta F)^2}{\gamma}\cdot\frac{\delta_S}{S_{H}\sqrt{N}},
\end{equation}
a factor of $\alpha (\Delta F)^2/\gamma$ larger than the small-inhomogeneity limit of Eq.~\ref{eq:ultimate_limit}. In this case, the measurement sensitivity could actually be improved by using states with smaller polarizability.

Equation~\ref{eq:homogeneous_ef}, in combination with the data shown in Fig. 2 of the main text, can be used to estimate the effects of field inhomogeneity and detection noise in our experiment. For example, when measuring the field several mm away from the MOT, the maximum $dS/dF$ occurs at $F\approx 0.5 \unit{V/cm}$. The polarizability of the $36s_{1/2}$ Rydberg state is $\alpha=5.2 \unit{MHz/(V/cm)^2}$. Under these conditions, linewidth is typically $\Gamma=4 \unit{MHz}$, signal/noise $\delta_s/S_o\approx 0.1$ and we make $N\approx 60$ measurements in the region of reasonably large $dS/dF$. The estimated measurement uncertainty under these conditions is therefore $\delta_F\approx 0.015 \unit {V/cm}$, a figure reasonably consistent with the measured repeatability of $0.04 \unit{V/cm}$. The resonance is not significantly broadened by field inhomogeneities, so measurement precision could potentially be improved by the use of larger fields or states with higher polarizability.

Close to the chip, inhomogeneous fields broaden the linewidth to $\Gamma\approx 20 \unit{MHz}$, and the longer duty cycle associated with loading atoms into the chip trap reduces the typical number of measurements to $N\approx 15$. Signal/noise is similar to the MOT release case, and maximum $dS/dF$ occurs at $F\approx 2 \unit{V/cm}$. The estimated measurement uncertainty under these conditions is $\delta_F\approx 0.04 \unit {V/cm}$. This estimate is smaller than both the observed day-to-day repeatability of $0.6 \unit{V/cm}$ and the intra-day repeatability of $0.15 \unit {V/cm}$. However, this model does not take into account any time-variation of the fields so the discrepancy is hardly surprising. The transition is broadened significantly even at zero field, so in this case measurement precision could be improved by using an excited state with lower polarizability.

\end{document}